\documentclass[useAMS,usenatbib,onecolumn]{mn2e}

\usepackage{graphicx}
\usepackage{natbib}
\usepackage{cite}

\bibpunct{(}{)}{;}{a}{}{,}
\usepackage{hyperref} 
\newcommand{\aap}{A\&A}

\newcommand{\aj}{AJ}

\newcommand{\apj}{ApJ}
\newcommand{\apjs}{ApJS}
\newcommand{\mnras}{MNRAS}
\newcommand{\araa}{ARA\&A}
\newcommand{\pasp}{PASP}

\newcommand{\nat}{Nature}

\newcommand{\na}{NewA}

\title[Jet-cloud interaction in 3C 381.]{Searching for evidence of jet-cloud interaction in radio galaxies. First results for 3C 381.}
\author[V. Reynaldi and C. Feinstein]{V. Reynaldi$^{1,2}$\thanks{E-mail:
vreynaldi@fcaglp.unlp.edu.ar (VR); cfeinstein@fcaglp.unlp.edu.ar (CF)} and C. Feinstein$^{1,2}$\footnotemark[1]\thanks{Based on observations obtained at the Gemini Observatory, which is operated by the Association of Universities for
    Research in Astronomy, Inc., under a cooperative agreement with the
    NSF on behalf of the Gemini partnership: the National Science
    Foundation (United States), the Science and Technology Facilities
    Council (United Kingdom), the National Research Council (Canada),
    CONICYT (Chile), the Australian Research Council (Australia),
    Minist\'{e}rio da Ci\^{e}ncia, Tecnologia e Inova\c{c}\~{a}o (Brazil)
    and Ministerio de Ciencia, Tecnolog\'{i}a e Innovaci\'{o}n Productiva
    (Argentina)}\thanks{Based on observations made with the NASA/ESA {\it Hubble Space Telescope}, which is operated byt he Association of Universities for Research in Astronomy, Inc, under NASA contract NAS 5-26555}\\
$^{1}$Facultad de Ciencias Astron\'{o}micas y Geof\'{i}sicas, UNLP, Paseo del Bosque s/n, La Plata 1900, Argentina\\
$^{2}$Instituto de Astrof\'{i}sica de La Plata, CONICET, Argentina}

\begin{document}

\date{}

\pagerange{\pageref{firstpage}--\pageref{lastpage}} \pubyear{}

\maketitle

\label{firstpage}

\begin{abstract}
We present results of {\it Gemini} spectroscopy and {\it Hubble Space Telescope} imaging of the 3C~381 radio galaxy. Possible ionising mechanisms for the Extended Emission-Line Region were studied through state-of-the-art diagnostic analysis employing line-ratios. Photoionisation from the central engine as well as mixed-medium photoionisation models fail in reproducing both the strengths and the behaviour of the highest-excitation lines, such as [Ne~{\sc v}]$\lambda3424$, He~{\sc ii}, and [O~{\sc iii}]$\lambda5007$, which are measured at very large distances from the AGN. Shock-ionisation models provide a better fit to the observation. Expanding shocks with velocities higher than 500~km~s$^{-1}$ are capable of reaching the observed intensity ratios for lines with different ionisation states and excitation degrees. This model also provide a direct explanation of the mechanical energy input needed to explain the high-velocity line-splitting observed in the velocity field.

\end{abstract}



\section{Introduction}

One of the most striking features found in radio glaxies (RG) has been the presence of large regions of ionised gas not only confined to central regions but also extending from tens to hundreds of parsecs from the host galaxies \citep*{fos89,car93,bin96,sol01}. These Extended Emission-Line Regions (EELR)\footnote{they are also known as Extended Narrow-Line Region (ENLR) when they are referred to Seyfert galaxies.} received their name for being located towards the outskirts of the galaxies, and far beyond the Narrow-Line Region (NLR), which is closer to the AGN \citep{pet97}. However, while the NLR has shown itself as a direct consequence of the nuclear activity \citep[][and references therein]{pet97}, the origin of the EELR's gas and the mechanisms that trigger its emission are still subject of debate \citep{sol02}.

These EELRs are characterised by emission of narrow forbidden lines of several species with a variety of ionisation states and excitation degrees, together with strong Hydrogen recombination lines \citep{car93,pet97}. Also, the shapes, intensities and morphology of this gas vary from object to object.

There already exist studies that could give some clues to the physics involved in the emission of these large regions. For example, \citet*{car87,cha87} showed that the UV continua of powerful radio sources at $z > 0.6$ are closely aligned with their radio sources. For lower redshifts and lower luminosity radio galaxies, alignments of the extended emission-line gas with the radio structures were also found \citep[see][]{baum89}. One possible explanation for these observations is that if it is assumed that the EELRs could be bright, large and unresolved from the host galaxy, the optical shape of the object would be dominated by the EELR in the case of optically poor spatially resolved radio galaxies. So, this is a strong indication that the radio-jet could be the triggering mechanism for the ionisation of these regions. 
    
With the higher resolution instruments, several studies were carried out with the aim of understanding what kind of mechanisms can modify the physical condition of the kpc-scale interstellar or intergalactic medium (ISM or IGM). Specifically, any ionizing mechanism shoud be able to explain the emission of radiation in a place with no apparent local source of ionising photons, where the stellar radiation could not account for the large amount of photons that are being observed. A very wide sample of objects, among them Seyfert and radio galaxies, have been imaged by HST and many of them show large amounts of highly ionized and disturbed gas on their outskirts. 

Two important issues concerning these extended ionised regions remain uncertain and they are being discussed: their origin and the mechanisms involved in their ionization. We have started to investigate the latter using the most acceptable theories to date: photoionisation, mixed matter \citep{bin96,bin97} and shock-ionisation \citep{dop95,dop96,all08,gro10}.

Photoionisation assumes that the UV photon flux produced by the AGN ionises the line-emitting gas; it is described by the dimensionless ionising parameter {\it U}, which accounts for radiation field dilution as the distance from the nucleus increases. More complex photoionisation models were performed by \citet{bin96,bin97} who introduced the presence of gas clouds with different optical depths, and different values for {\it U}. On the other hand, shock-ionisation is triggered by shock waves driven by the radio jets. Those shock waves compress, heat, and accelerate the IGM, thus causing it to produce emission. \citet{dop95,dop96} and later \citet{all08} and \citet{gro10} developed MHD models for shock-ionisation, for both low and high shock velocities. When high speeds are considered, a radiative precursor region is created in front of the shock wave by the most energetic photons. As a consequence, the emission region gets bigger (it is composed by the precursor and the hot shocked gas) and the spectrum shows a mixture of ionisation and excitation states.

Both scenarios have been tested in a wide sample of active galaxies. \citet{ca95a,ca95b,cap96,cap97} and \citet{win97} were the first to show that shock-ionisation best explains the optical emission in the NLR in nearby Seyfert galaxies (Mrk 3, Mrk 6, Mrk 573, NGC 1068, NGC 4151 and NGC 7319). Recently, \citet{ros10a,ros10b} have shown that this mechanism also occurs in NGC 5929 and SDSS J1517+3353. In the case of powerful 3CR radio sources, \citet{fei99,fei02,mar00,sol01,sol02,sol03,til05,chr06,har10} showed that this interaction can be found in the EELRs of 3C~299, 3C~244.1, 3C~171, 3C~34, 3C~330, 3C~352, 3C~435A, 3C~265, 3C~196, 3C~277.3, and also in PKS~2250-41. In contrast, \citet{rob00,sol04} have proved that 3C 321 and the powerful radio galaxies 0850-206 and 1303+091 are photoionised by the AGN.

\subsection{The galaxy 3C 381}

3C~381 is a FR II double-lobed radio source. It has a flux density of 16.6 Jy at 178 MHz \citep*{lai83}. The radio structures are located in an approximate North-South direction \citep[see figs.~ 21-22 from][]{lea91}, the radio-axis being defined by the line that connects the radio core with the two hotspots. 3C~381 is a large $\sim 74$ arcsec radio source, the south lobe is larger than the north one \citep[40.5 and 33.4 arcsec, respectively][]{lea91}; its inner edge is more separated from the central source than the north lobe is. None the less, a fairly symmetric rotational pattern can be distinguished near the centre\footnote{See the DRAGNs Web site, http://www.jb.man.ac.uk/atlas/, which is maintained at the Jodrell Bank Observatory by J.P. Leahy, A.H. Bridle, and R.G. Strom.} despite the separation.

The galaxy was studied at optical wavelengths by \citet*{car95}, who showed the presence of extended [O {\sc iii}]$\lambda5007$ emission. They characterized the object as one with the largest high-ionisation emission-line regions amongst the low-redshift (z=0.1605) 3CR sources, which was later confirmed by HST/WFPC2 data (Fig.~\ref{fig-1}). The optical and radio features are not aligned. However, near the centre both radio lobes are rotationally symmetric distorted and misaligned with respect to the radio axis (defined by the passage of the jet) in the same way as the [O {\sc iii}]$\lambda5007$ extended emission is. Moreover, the north-west optical elongation reaches the south extreme of the north lobe \citep[see fig. 54 from][]{pri08}, but there is also an additional much more extended [O {\sc iii}]$\lambda5007$ filament 25 arcsec to the north \citep[fig. 5 from][]{car95} that seems to be either embedded in or superimposed on the north lobe, in a direction approximately coincidental with the radio axis. The galaxy was classified as a High Excitation Galaxy (HEG) based on its strong nuclear spectrum and its [O {\sc iii}]$\lambda5007$ luminosity \citep{but09,but10}.

As all galaxies in the 3CRR catalogue, 3C~381 is located away from the Galactic plane \citep{lai83}. The extinction is low enough ($E_{(B-V)}= 0.053$~mag; NASA/IPAC Extragalactic Database) that the Galactic reddening correction is negligible. Throughout this paper, we assume $H_0 =73$~km~s$^{-1}$~Mpc$^{-1}$,  $\Omega_{matter}=0.27, \Omega_{vacuum}=0.73$. Hence, at a redshift of $z=0.1605$, 3C 381 is at a distance of 768~Mpc with a projected linear scale of 3.2~kpc~arcsec$^{-1}$.

We are going to investigate the emission coming from the EELR of 3C~381 by using long-slit GMOS/Gemini spectroscopy with the aim of determining which of the ionising mechanisms mentioned above best explain the ionisation state of this large-scale gas.
%
%

\section{Observations}

\subsection{HST/WFPC2 imaging}

3C~381 radio galaxy was observed with {\it HST}/WFPC2 as part of the 3CR Snapshot Survey (PI: Sparks). The log of observations is summarized on Table \ref{tbl-hst}. Two broad-band F702W images were taken in 1995 February (140s each one) which account for continuum emission from the host galaxy. Other two narrow-band ramp filter FR533N33 (300s) images were taken in 1995 August. This filter was employed to find ionised gas surrounding the galaxy. The WFPC2 scales are 0.0455~arcsec~px$^{-1}$ (146~pc~px$^{-1}$) for PC mode, and 0.0996~arcsec~px$^{-1}$ (319~pc~ px$^{-1}$) for WF mode.

The F702W filter is close to Cousins's R; these images were reduced as described in \citet{mar99} and then they were registered and added in order to produce one final image with higher signal-to-noise ratio (S/N). Data were reduced using the {\sc stsdas} package within {\sc iraf}, which includes the standard WFPC2 pipeline processing and cosmic ray removal.  Given the redshift of the galaxy and its position on the CCD, the ramp filter produces an [O {\sc iii}]$\lambda5007$-dominated image (Fig.~\ref{fig-1}, right panel). These structures are located at PA $155^\circ$, they extend up to 2.2 arcsec to the North-West and South-East from the nucleus respectively, and they constitute the targets for our spectroscopic analysis.

As the HST images are flux calibrated data, photometry was done in order to calibrate the exposure time needed to have a good S/N ratio for the spectroscopy. This procedure was performed by measuring the [O {\sc iii}]$\lambda5007$ emission and then its flux was scaled to the oxygen line in a quasar-like spectrum, which is the most similar template available. Both from the images and this analysis, it is clear that the nebular structures are very extended and they are bright enough to stand out over the stellar luminosity. However, our main interest is focused on the regions where the stellar component already vanishes, or it is so faint that there is no measurable contribution in the spectra.

\subsection{GMOS spectroscopy}

The long-slit spectrum was obtained in 2005 April using {\it Gemini's Multi-Object Spectrograph} (GMOS) at the 8.1~m {\it Gemini North} telescope as part of the program GN-2005A-Q-37 (PI: Feinstein; see Table~\ref{tbl-gem} for the log of observations). The instrument was set up with a 0.5 arcsec width slit, and the B600-G5303 grating  R$\sim$1700 centred at about 4500 \AA, which yields a resolution of 0.9 \AA px$^{-1}$.  Only one exposure of 2400s was taken with 2$\times$2 binned CCD, implying a spatial resolution of 0.1454~arcsec~px$^{-1}$. The slit was placed at a position angle (PA) of $155^\circ$, along the optical-UV elongation.

Data were reduced by using the {\sc gemini-gmos} package reduction tasks within {\sc iraf}. We followed the usual steps of bias subtraction, flat-field correction, wavelength calibration, and sky subtraction. We performed cosmic ray rejection from the 2-dimensional spectrum by using two different tasks: {\it gscrrej} from {\sc gemini} package and the Laplacian Cosmic Ray Identification \citep{dok01}, but we found some loss of information after the tasks were (separately) carried out. For these reasons, we decided not to apply this correction to our spectrum. However, the presence of a cosmic ray hit over a given profile in the extracted 1-dimensional spectrum is clearly detected because of its shape, intensity, and width relative to the line itself, so cosmic ray contamination was easily removed from our measurements.

The spectrum covers $\sim$2700 \AA$\ $ in wavelength, from 3800 \AA$\ $ to 5800 \AA. It includes emission-lines of several species, from the bluest [Ne {\sc v}]$\lambda3424$ to the reddest [O {\sc iii}]$\lambda5007$, the latter being the most important feature in the spectrum (Fig.~\ref{fig-2}). The set of identified lines is listed in Table \ref{tbl-lines}, together with line fluxes relative to that of H$\beta$ for each selected position. All the locations are referred to the galactic centre,  which was morphologically determined both for the spectra as for the HST imaging. Since 3C 381 is an elliptical galaxy, and it is well fitted by a De Vaucouleurs profile, the centre can be determined very accurately for the HST continuum image. Concerning Gemini spectrum, the centre was located as the maximum intensity point within the stellar continuum. This point matches the maximum of the De Vaucouleurs profile.

Since Gemini spectrum has a resolution of 0.1454~arcsec~px$^{-1}$, it gives a linear scale of about 4654~pc~px$^{-1}$, without projection correction. Fig.~\ref{fig-2} covers the range from H$\beta$ to [O {\sc iii}]$\lambda5007$; the intensities of these lines are shown in gray scale; the contours have been plotted with the aim of emphasizing the line's intrinsic shapes. 

%
%

\section{Results}

\subsection{HST/WFPC2 imaging}

\begin{figure}
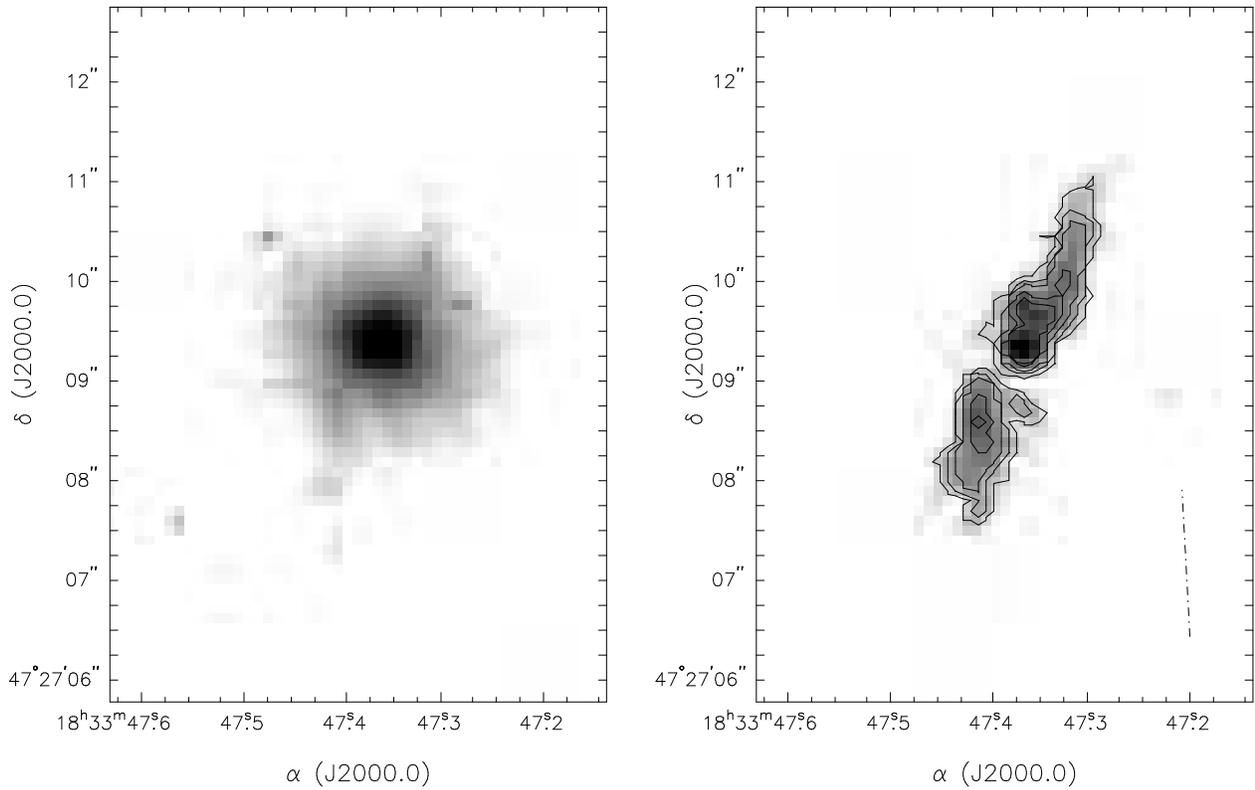

  \centering
  \hfill\includegraphics[angle=270,width=.45\textwidth]{fig-1-left.eps}~\hfill\includegraphics[angle=270,width=.45\textwidth]{fig-1-right.eps}\hfill~
\caption{HST/WFPC2 images of 3C~381. Left: The stellar component accounting for the elliptical structure. Broad band filter image. Right: [O {\sc iii}]$\lambda5007$ -dominated image shown in grey-scale and contour map. The contours levels are 5\%, 7\%, 10\%, 18\% and 22\% of the maximun intensity at the nucleus position. The radio axis direction \citep[PA=$4^\circ$,][]{koff96} is also shown at the bottom right of this panel. \label{fig-1}}
\end{figure}

\begin{figure}
\begin{center}
\includegraphics[scale=0.9,angle=0]{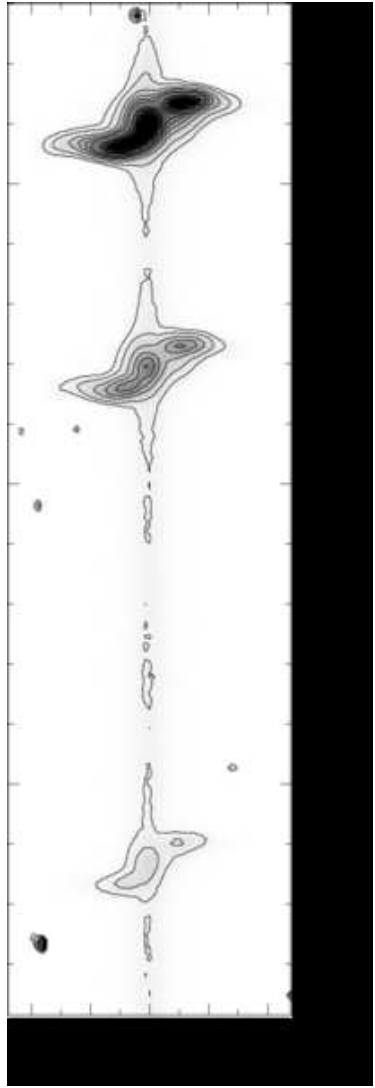}
\caption{Two-dimensional Gemini spectrum, from H$\beta$ to [O{\sc iii}]$\lambda5007$. The ordinate scale indicates the SE (negative) and NW (positive). The contour levels are shown only to enphasise the grey-scale contrast. \label{fig-2}}
\end{center}
\end{figure}

The galaxy shows a completely different morphology at broad band compared with the narrow band. While the first shows the stellar component dominating the flux, the narrow band filters are sensitive to the ionised gas. Its elliptical nature is clearly seen in broad-band filter (Fig.~\ref{fig-1}, left panel). The elliptical structure is homogeneous and it is well fitted by a De Vaucouleurs ($r^{1/4}$) surface brightness profile. It vanishes below the noise in the HST images at 1.7 arcsec from the nucleus, the stellar component being negligible as from this point. 

On the other hand, the HST/WFPC2 ramp filter image shows the nebular emission dominated by [O {\sc iii}]$\lambda5007$. These structures are elongated and they vanish in the outskirts of the host galaxy, giving a total extension from one end to the other of about 4.5 arcsec at a position angle of $155^\circ$ (counterclockwise from the North). The shapes of both features are quite different. The NW EELR is narrower than that of the SE; it has its maximal emission concentrated over its inner region from where it adopts an {\it S} shape towards its outer limit, reaching the south extreme of the northern radio lobe. The SE EELR, in contrast, is clearly detached from the central object. The peak of its emission is located at 1.1 arcsec from the nucleus (Fig.~\ref{fig-1}, right panel).
 
\subsection{GMOS spectroscopy}

The high-excitation nature of 3C~381 established on the basis of its nuclear spectra is emphasized with our data, where high- and low-excitation lines coexist in a region that extends up to 3.2 arcsec (10~kpc) from the nucleus in each direction. However, the [O {\sc iii}]$\lambda5007$ emission is very intense, so it is measurable at a larger distance from the nucleus with a higher S/N than any other species. It extends up to 4.8 arcsec to the NW and  7.3 arcsec to the SE.

The angular size of the entire EELR measured by the [O {\sc iii}]$\lambda5007$ line reaches 12 arcsec (38.4~kpc), an extension of almost three times what the HST had detected, which indicates the existence of very low surface brightness structures towards the outer edges of the EELR.

The line shapes in the long-slit spectrum show a strong core over the central region, with a blueshifted wing to the NW and a redshifted wing to the SE (Fig.~\ref{fig-2}). Given the complexity of the lines, we have extracted one 1-dimensional spectrum for each spatial pixel across the slit to avoid the loss of spatial information. Once the extraction was performed, we found a complex spectral structure, where the line profiles can be decomposed into two or even three (in the case of [O {\sc iii}]$\lambda\lambda4959,5007$) Gaussian components.

The Gaussian decomposition was performed by using the {\it ngaussfit} task from the {\sc stsdas} package and the {\it splot} task from the {\sc noao} package within {\sc iraf}. The total flux together with its rms error were obtained with {\it ngaussfit} and then checked with {\it splot}. The {\it ngaussfit} task works by iteration based upon a set of initial conditions on its three free parameters: central wavelength, intensity at the peak and {\sc fwhm} for each component, in addition to the baseline continuum. We set initial conditions for the central wavelengths and line intensities; the {\sc fwhm}s were left to converge by themselves. Finally, the initial values we proposed did not change significantly over the iterations and all the profiles reached their best fit with narrow ({\sc fwhm} $<800$~km~s$^{-1}$) components for both forbidden and permitted lines. The line intensities decrease from the nucleus to the outskirts for the north-west EELR, but there is a sudden increase towards the south-eastern region at around 1 arcsec ($\sim 3.2$~kpc). This behaviour was detected in the most intense lines as well as in the weakest ones. It reveals that the emission increase towards the SE in the [O {\sc iii}]$\lambda5007$ contour-map (Fig.~\ref{fig-1}, right panel) is common to the whole EELR SE-gas.

Strong line-splitting was found in oxygen forbidden lines but also in hydrogen recombination lines. Although 3C~381 has been considered as a Broad-Line Radio Galaxy (BLRG) on the basis of a weak broad component in H$\alpha$ \citep{gra78}, no broad component was needed to reconstruct the profile in any Hydrogen line. Even when H$\alpha$ lies outside the wavelength range we cover, we are able to say that 3C~381 has a Seyfert II-like spectrum. It has been also confirmed not only by X-ray nuclear data taken with {\it Chandra} \citep*[][and references therein]{har09}, but also by recent optical nuclear data taken with {\it Telescopio Nazionale Galileo} (TNG) \citep{but09}, which shows no evidence of such a broad component. For these reasons we assume 3C~381 is a Narrow Line Radio Galaxy (NLRG), which is in keeping with its lack of a quasar-like core. 

In addition to this line-splitting, the strongly distorted [O {\sc iii}]$\lambda5007$ emission-line profile also has a broad underlying component, mainly manifested inside the inner $\pm$3 arcsec ($\pm10$~kpc). The broad wings in [O {\sc iii}]$\lambda5007$ (which are also detected in [O {\sc iii}]$\lambda4959$) make it necessary to use more than a single narrow gaussian fit. The narrow component can be easily followed throughout the spectra, and its parameters are measured with high accuracy. On the other hand, the errors associated with the broad component parameters are high, specially the {\sc fwhm}. The measurement of this component is much more difficult than that of the narrower one, but its underlying presence is not arguable.

\subsubsection{Kinematics of the emitting gas}
\label{kinematics}

\begin{figure}
\begin{center}
\includegraphics[scale=1,angle=0]{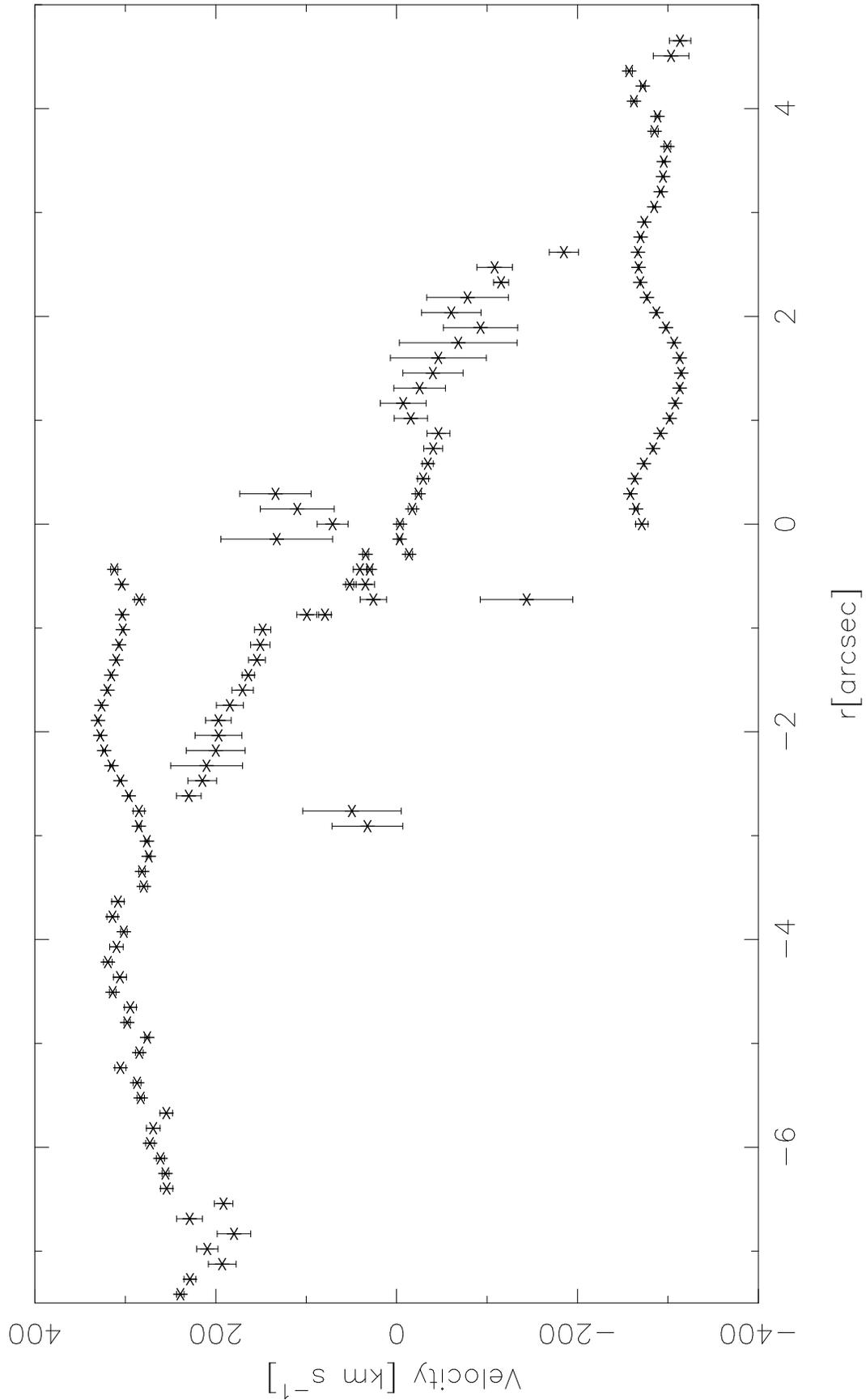}
\caption{Velocity field computed from the components that form the [O {\sc iii}]$\lambda5007$ emission-line profile. As in Fig.~\ref{fig-2}, SE is plotted as negative coordinates and NW as positive.\label{fig-3}}
\end{center}
\end{figure}

Due to its brightness, the [O {\sc iii}]$\lambda5007$ emission-line was used to compute the velocity field relative to the host galaxy systemic velocity (Fig.~\ref{fig-3}). A fairly symmetric distribution to each side of the nucleus is distinguished as an organized and collective pattern, coming from a gas component probably related to a disk. This rotation curve tends to stabilize around 300~km~s$^{-1}$ at $\sim$1.5 arcsec ($\sim 5$~kpc) from the galactic centre which is consistent with gravitational motion within the galactic potential \citep*{tad89}. However, there are also high velocity components superimposed on that curve which follow neither ordered nor collective behaviour. Such a velocity shift cannot be explained under the gravitational hypothesis. The highest velocity components are located within the inner $\pm$3 arcsec ($\pm10$~kpc); they are separated from the rotation curve by $\pm300$~km~s$^{-1}$ approximately. They deserve special attention because those disturbed kinematic components are also the broader ones (Fig.~\ref{fig-3}).

The high-velocity displacements among the kinematic components could be related to disturbing processes triggered by the radio-emitting components in different scenarios. Evidence of such processes has been found not only in radio galaxies where the radio components overlap the line-emiting gas, but also in radio sources where the radio structures are found far beyond the innermost EELR \citep{sol01}.
The same happens in 3C~381; the radio and optical features have different scales, excepting the outer edge of the NW EELR that reaches the south extreme of the northern lobe, in addition to the other ionised filament, detected in the same location of the north lobe \citet{car95}. 3C 381 is one of these galaxies where the interaction between radio and optical features can be identified even when these components are not overlapped.

The kinematic disturbed components we find are strong indicators that this gas is undergoing a violent disturbing process, very probably triggered by the radio structures. So, any modelling of the undergoing emission process must also account for the energy that is accelerating this component of gas.

%
%

\section{Discussion}

To understand and identify the main ionising mechanism of the EELR, we test the line-ratios from the 3C~ 381 spectrum with the results predicted from modelling of several mechanisms that could ionise the interstellar gas. In this process we have to identify the main source of energy of the line-emission, which could be the AGN (e.g. AGN-photoionisation, mixed-matter models) or a local phenomena (eg. shock-ionisation). Most of these mechanisms are already modeled with predicted line-ratios that can be easily contrasted with the observed data.   

A direct way to model the problem of the ionising photon budget is to calculate the dimensionless ionisation parameter {\it U} through the empirical relationship \citep{pen90} over the entire nebular region. 

\begin{displaymath}
U = -2.74 -log \left( \frac{[OII]\lambda 3727}{[OIII]\lambda 5007} \right)
\end{displaymath}

This is the simplest way to understand if the UV photons coming from the AGN could be responsible for these ionised regions of extended gas. This relation was obtained under the assumption of solar abundances and gas density of 100~cm$^{-3}$. It covers all possible configurations applicable to AGN, with spectral indices $\alpha$ (${\Phi}_{\nu}\sim\ {\nu}^{+\alpha}$) varying in the range of $-2<\alpha<-1$ and black body temperatures in the range of $10^5\ K<T<2.1\times 10^7\ K$. The ionisation parameter is used to find an estimation for the amount of ionising photons emitted per time unit by the AGN ($Q_0$), for which the ambient gas has to be known ($Q_0=U4\pi r^2cn_e$). \citet*{rob02} performed long-slit spectroscopy and determined that at 1.4~kpc from the nucleus (PA=$155^\circ$) the density has a value of 370~cm$^{-3}$.

We measured $Q_0$ by using this local density together with {\it U} calculated in the same location where the density was obtained. At this point (1.4~kpc), $U=9.35\times 10^{-2}$. As our long-slit spectrum allows the spatial measurement of [O {\sc ii}]$\lambda3727$ and [O {\sc iii}]$\lambda5007$, we can compute the line-ratio and, under the assumptions of Penston's relationship, trace the spatial evolution of {\it U} (Fig.4). Even if the density changes as a function of position, $Q_0$ is supposed to be constant. So, we obtain $Q_0=2.4\times 10^{56}$~photons~s$^{-1}$. Both $Q_0$ and {\it U} must be considered as a minimum because the inclination is unknown, and it has not been corrected.

This $Q_0$ is comparable to that of 3C~273, the most powerful radio-loud quasar known at low redshift ($Q_0=3.8\times 10^{56}$~photons~s$^{-1}$, \citet{rob00}\footnote{their published value is $Q_0=3\times 10^{55}$~photons~s$^{-1}$~ster$^{-1}$}). \citet{har09} have shown that what they called X-ray 'accretion related' nuclear luminosity (absorption corrected) is a good indicator of the AGN power. The X-ray luminosity of 3C~381 is almost an order of magnitude lower that that of quasars \citep[see fig.~13 from][]{har09} (and the [O {\sc iii}]$\lambda5007$ luminosity too). So it is very likely that the $Q_0$ previously calculated is a very overestimated measurement of the ionizing power in 3C~381.

\begin{figure}
\begin{center}
\includegraphics[scale=0.9,angle=0]{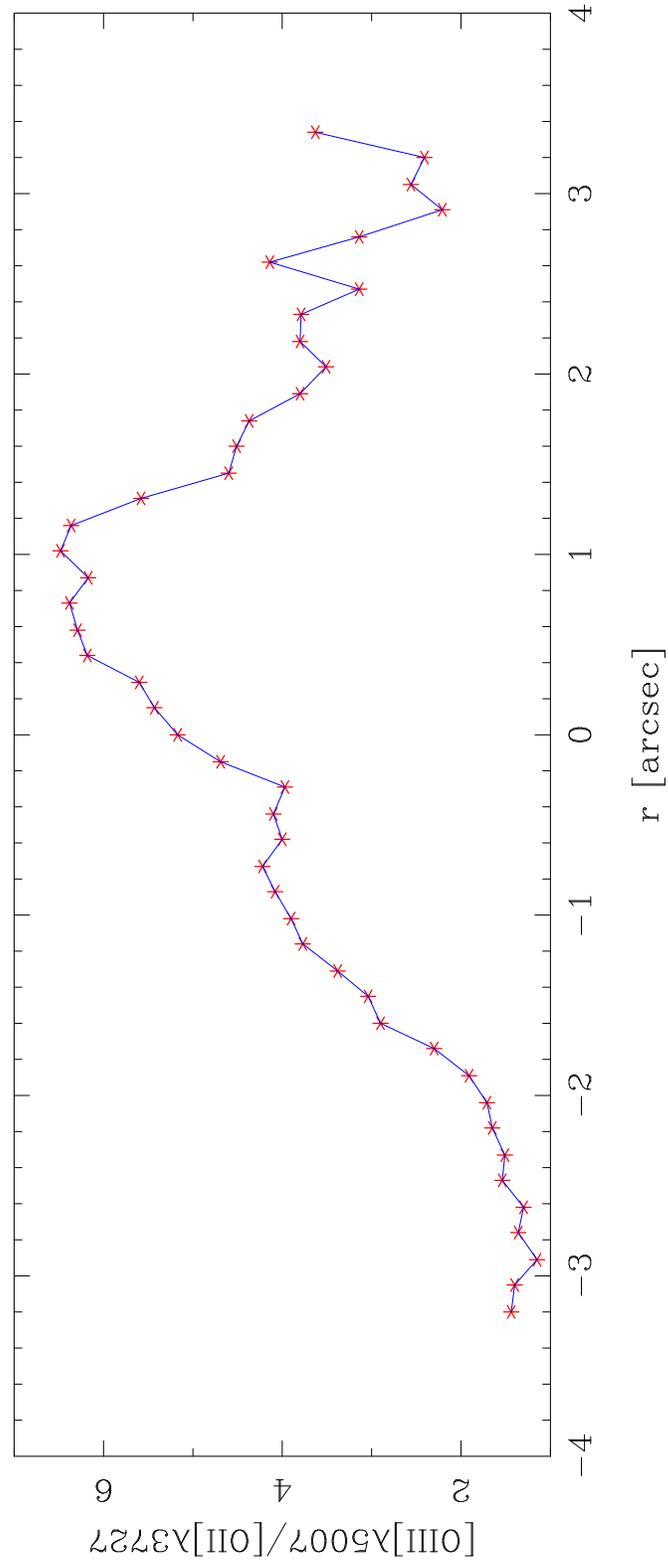}
\caption{Spatial variation of the ionisation parameter {\it U} expressed through the oxygen ratio \citep{pen90}.As in Fig.~\ref{fig-2}, SE is - and NW is +. Note that {\it U} does not follow the $r^{-2}$ geometrical dilution but grows up at $\sim$1 arcsec towards the SE. Every emission-line in 3C 381's spectrum increases its intensity at the same location. \label{fig-4}}
\end{center}
\end{figure}

In addition, other problem arises concerning {\it U}. We expect that {\it U} evolves as $U\sim n_e^{-1}r^{-2}$. So it would be sensible to the geometrical dilution of the central ionising radiation field, and the changes in the density. Fig.~\ref{fig-4} reflects the behaviour of {\it U}, through [O {\sc iii}]$\lambda5007$/[O {\sc ii}]$\lambda3727$, under the assumptions of constant density \citep{pen90}. It can be clearly seen that {\it U} does not follow the $r^{-2}$ dilution over a large distance. In contrast, {\it U} increases towards the south-eastern EELR and, even more, we must highlight that every emission-line in the spectra increases its intensity towards the SE (at $\sim 1$ arcsec). In order to produce such an increase in the emission (as {\it U} shows), the density should decrease faster than $r^{-2}$ over a kpc-scale distance, which seems to be unlikely. It is worth noting that the [O {\sc iii}]$\lambda5007$/[O {\sc ii}]$\lambda3727$ line-ratio will significantly change its predicted behaviour (with respect to the single ionising source) if there is other off-nuclear source of ionising photons, such as the {\it local} radiation field created by {\it local} radiative shocks. We will discuss it later.

Therefore, from the overestimated ionising budget and the observed behaviour of the  [O {\sc iii}]$\lambda5007$ and [O {\sc ii}]$\lambda3727$ line-ratio, we think that it is very unlikely that a simple AGN-photoionisation could be the dominant ionising mechanism.

Then, a more complex photoionisation model is required. A clever approach  was developed by \citet{bin96} doing a more sophisticated modelling of the way in which a clumpy medium absorbs the impining radiation field. In this mixed-matter model \citep{bin96,bin97}, the EELR is composed by two different cloudy systems. The {\it matter-bounded} (MB) clouds, which are the optically-thin lower-density cloudlets, absorb a fraction of the incoming radiation. Since the highest-energy photons are absorbed here, this is the place where the highest-excitation lines are formed. The {\it ionisation bounded} (IB) clouds absorb the diluted ionising field that exits the first component. The lowest-excitation lines in the spectrum are formed in this optically-thick cloudlets. In this clumpy system, the radiation field emitted by the central surce is partially absorbed by the MB clouds, and once it has been leaked, it is completely absorbed by the IB clouds due to their higher density. The extension of the MB component is defined by the availability of He$^+$ ionising photons \citep{bin96}, which in turn are manifested by the presence of He{\sc ii} recombination lines \citep{ost89}.

However, the He {\sc ii}$\lambda4686$ emission 3C 381 extends up to 2.5 arcsec ($\sim 8$~kpc) over each EELR, without the strong variation that the model proposes. The line intensity at the end of each EELR drops to $\sim 10$\% of its maximum value at the galactic centre, while the intensity of [O {\sc iii}]$\lambda5007$, the main feature in the spectrum, drops to less than $\sim 5$\% of its maximum value in the same spatial extent. In fact, the extent to which the line is measured is in itself a problem, since it sets the spatial limit for the MB clouds. Consequently, there is no mixed-material inside the inner 2.5 arcsec ($\sim 8$~kpc); therefore, there applies again the hypothesis proposed by the simple AGN-photoionization model discussed before, which fails in reproducing the scenario in 3C~381's environment.

\begin{figure}
\includegraphics[angle=270,scale=1]{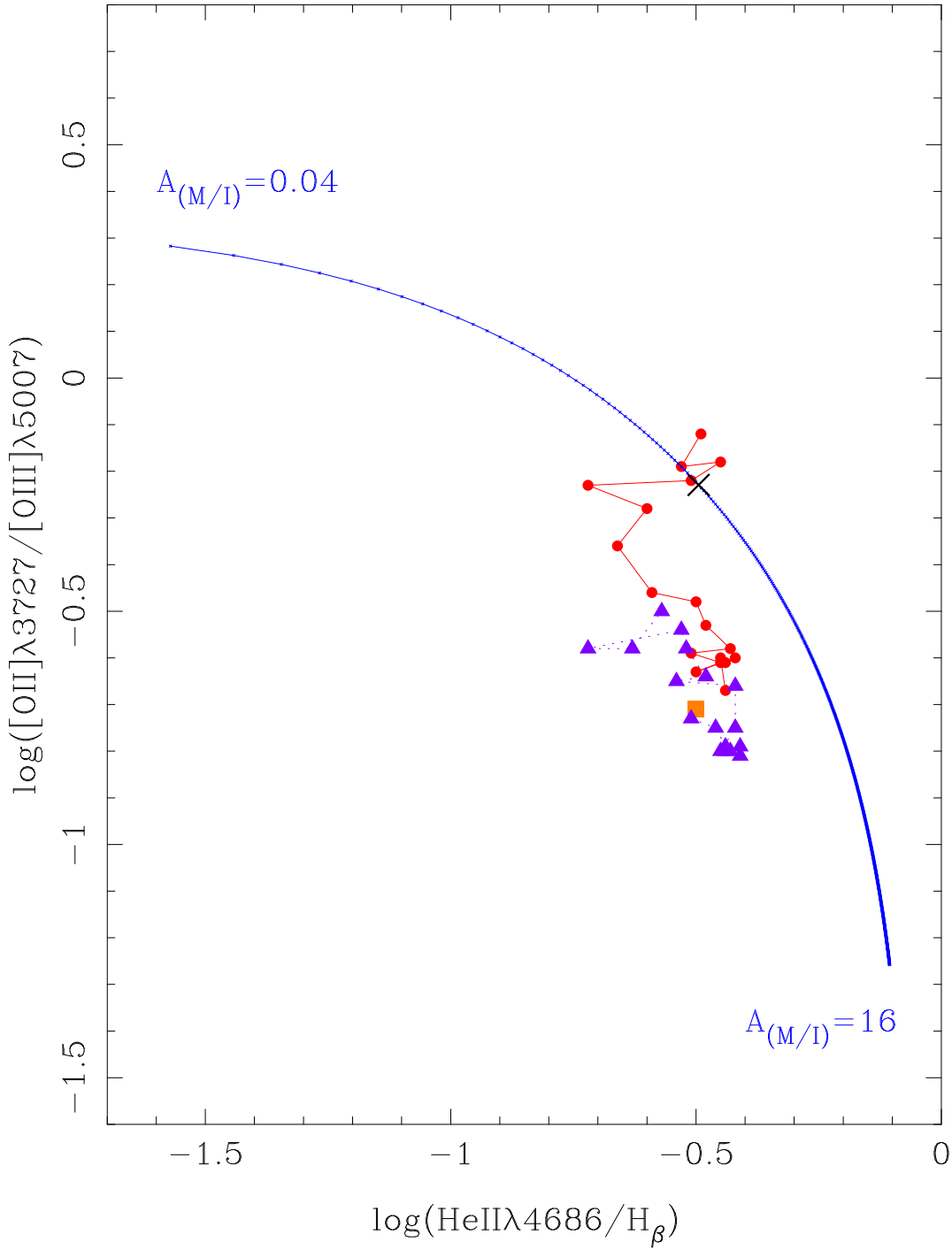}
\caption{Continuous line: mixed medium photoionisation model from \citet{bin96}. The black cross shows the position where $A_{M/I}=1$. Its extreme values are shown at the end of the sequence. Each EELR was plotted separately, the data are connected to follow their trend with respect to the $A_{M/I}$ sequence. The NW measurements were plotted as dots, and those from the SE as triangles; the central spectrum is shown by a square.\label{fig-5}}
\end{figure}

The modelling is implemented by diagnostic diagrams parameterised by $A_{M/I}$ (defined as the solid angle subtended by the MB clouds with respect to that of IB clouds) which varies as the He {\sc ii}/H$\beta$ line-ratio does. The diagram plotted in Fig.~\ref{fig-5} uses the oxygen ratio [O {\sc ii}]$\lambda3727$/[O {\sc iii}]$\lambda5007$ against He{\sc ii}/H$\beta$, which became an excitation axis given its relation with $A_{M/I}$. It is worthwhile comparing this diagram with fig. 7 from \citet{bin96} to ensure that the observed trend actually responds to changes in {\it U} rather than $A_{M/I}$ \citep[see also][for the same kind of analysis on Mrk 573]{ferr99}. Each EELR was plotted separately from the other, connecting the measurements to follow their trend with respect to the $A_{M/I}$ sequence. The NW measurements were plotted as dots, and those from the SE as triangles; the central spectrum is shown by a square. 

It is very unlikely that photoionisation models could account for the whole observations. Even when there should be a contribution from photoionisation, neither the observed increment in {\it U} toward the SE nor the line-ratios can be reproduced by these kind of models.

A plausible explanation for an increasing ionisation parameter (note that {\it U} increases because [O {\sc ii}]$\lambda3727$ increases as well as [O {\sc iii}]$\lambda5007$ does) is the shock-ionisation model, since either a local source of ionising photons is needed or the AGN has improved its photoionising power as a consequence of the disturbing processes undergone by the EELR material \citep{sol03}. This mechanism, also known as jet-cloud interaction, might provide a local ionising field if the expanding shock waves are fast enough, but such a scenario should be manifested not only in line-intensities but also in the kinematics of the region. 

When a shock wave is capable of ionising the medium where it propagates, the emission of radiation usually comes from the recombination region behind the shock front. But the fastest shocks ($v>170$~km~s$^{-1}$) create UV photons that can diffuse the jet both upstream and downstream, changing the ionisation conditions of the entire region. Photons travelling upstream create an H{\sc ii}-like region (known as the precursor) ahead the shock, but also the post-shock gas is affected by photons travelling downstream. As a consequence, the line-emitting region gets an important contribution from the precursor: it is 10\% to 20\% brighter than the shock in H$\beta$ \citep{dop95,dop96,all08}. 

\begin{figure}
\includegraphics[angle=0,scale=0.9]{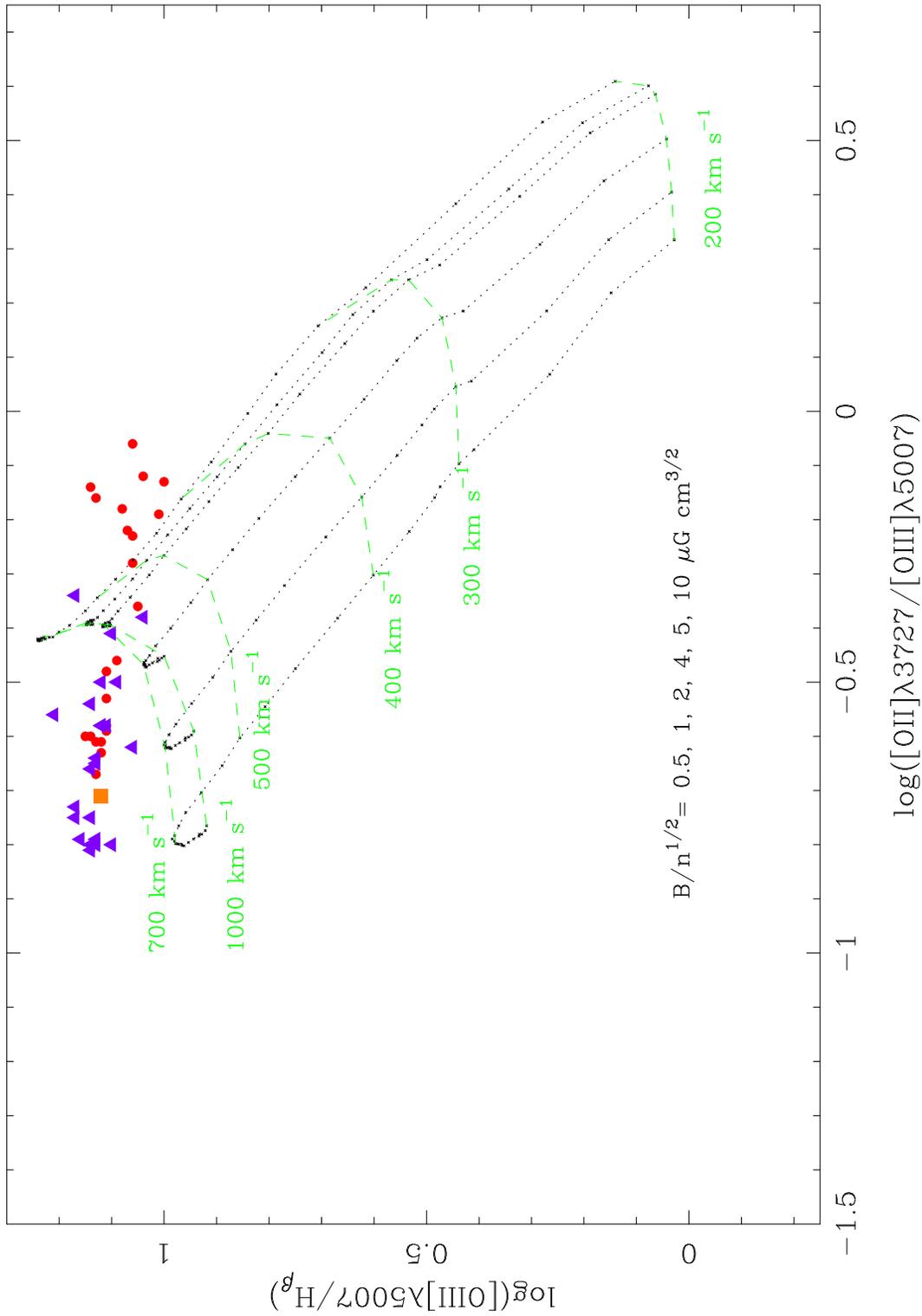}
\caption{Dotted lines: shock-ionisation + precursor models from \citet{all08}, one line per magnetic parameter ($B/n^{1/2}$) value. Shock velocities are drawn as dashed lines. See Fig.~\ref{fig-5} for data reference.\label{fig-6}}
\end{figure}

\begin{figure}
\includegraphics[angle=270,scale=1]{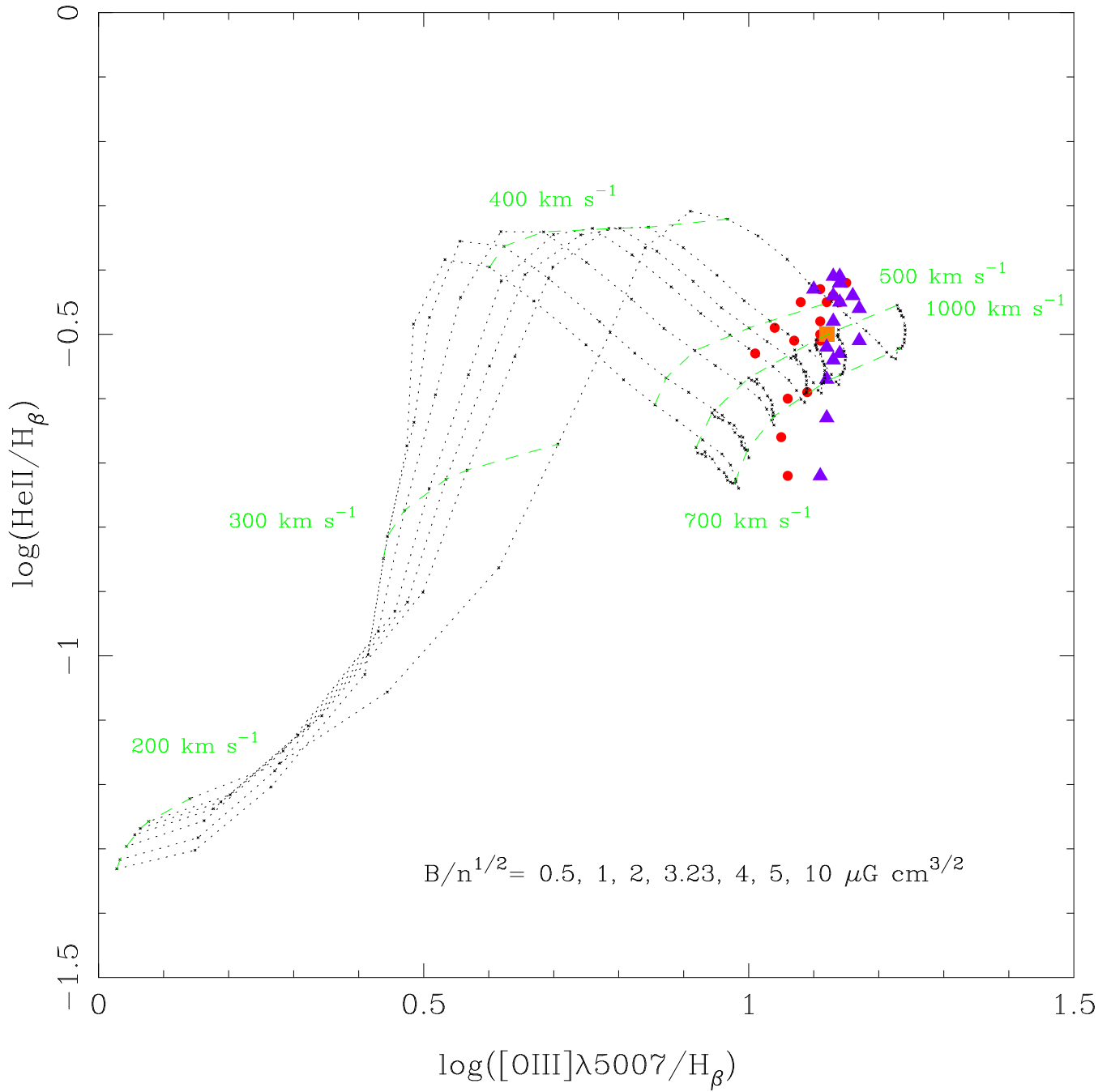}
\caption{This plot shows the relation between two excitation axes, [O {\sc iii}]$\lambda5007$/H$\beta$, and the other one implemented by \citet{bin96} He {\sc ii}/H$\beta$. See Fig.~\ref{fig-6} for model references. \label{fig-7}}
\end{figure}

\begin{figure}
\includegraphics[angle=270,scale=1]{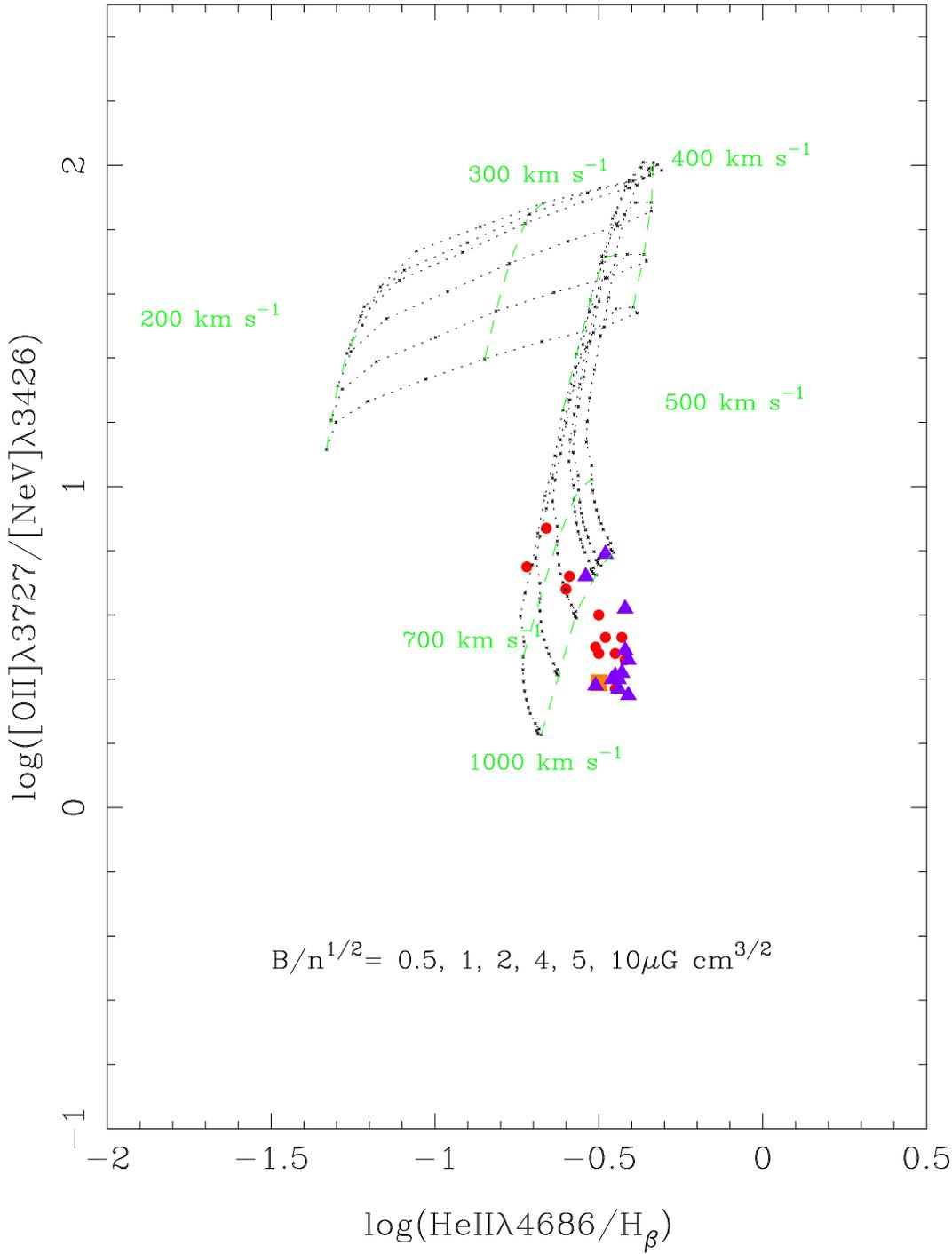}
\caption{The high-excitation lines [Ne {\sc v}]$\lambda3424$ and He {\sc ii} are combined with the lower excited [O {\sc ii}]$\lambda3727$ and H$\beta$. The shock-ionisation models can reproduce the observations when the highest velocities are considered.\label{fig-8}}
\end{figure}

\begin{figure}
\includegraphics[angle=270,scale=1]{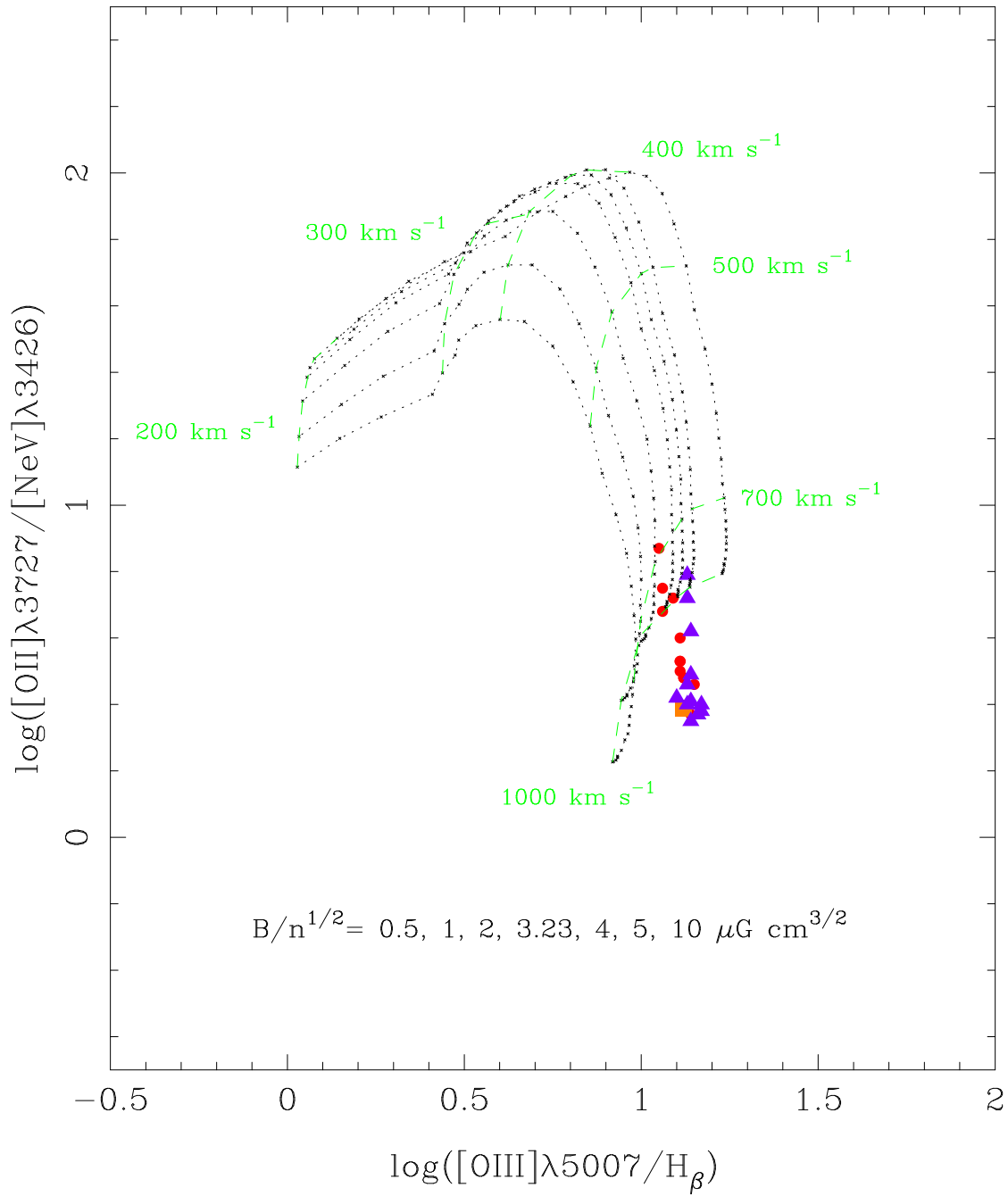}
\caption{The same than Fig.~\ref{fig-8} with [O {\sc iii}]$\lambda5007$/H$\beta$ as the excitation axis. See Fig.~\ref{fig-6} for model references. \label{fig-9}}
\end{figure}

\begin{figure}
\includegraphics[angle=0,scale=0.9]{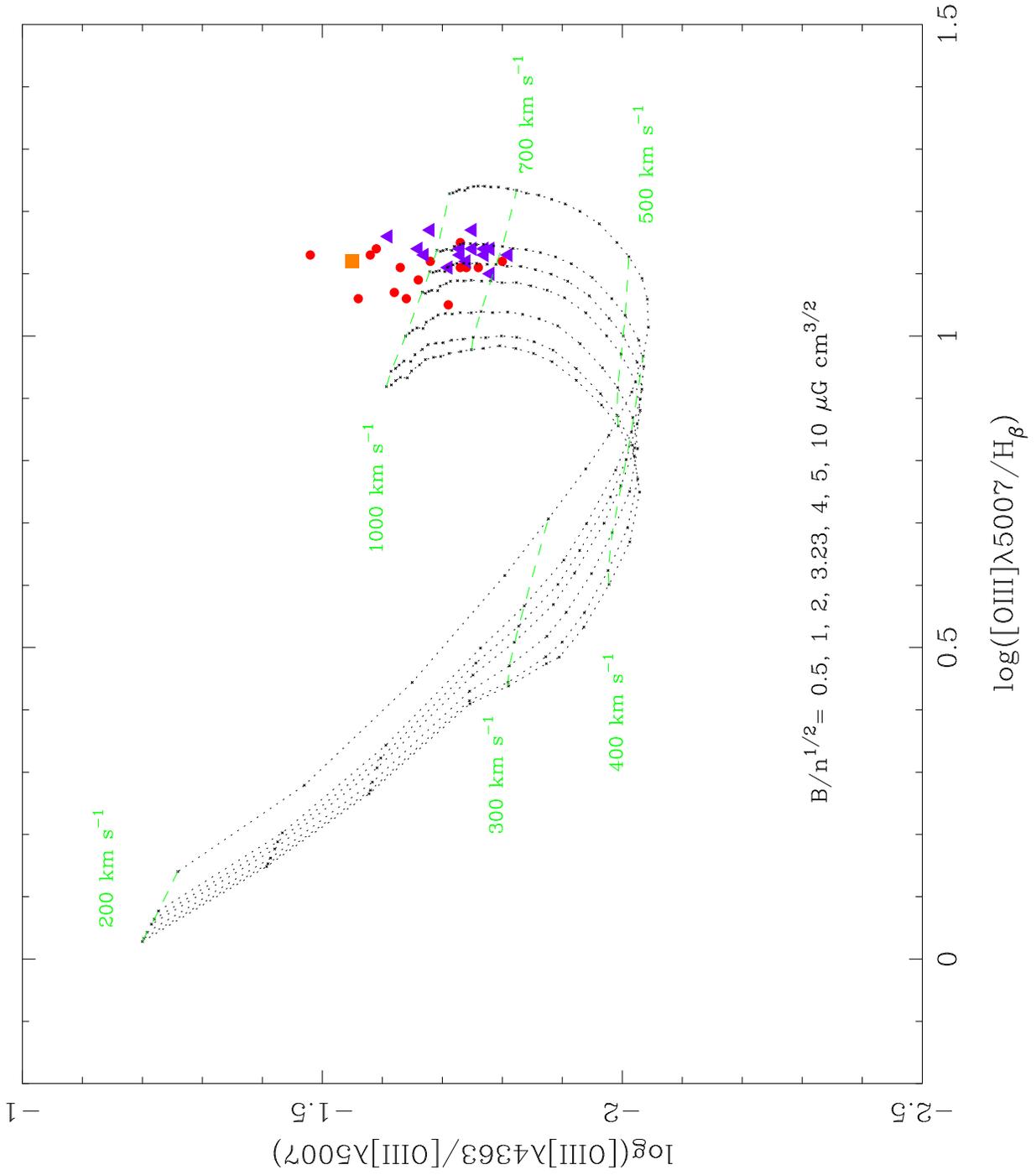}
\caption{The temperature-sensitive line ratio [O{\sc iii}]$\lambda4363$/[O{\sc iii}]$\lambda5007$ against [O{\sc iii}]$\lambda5007$/H$\beta$. Despite the {\it temperature problem}, the diagram shares the same characteristic of the previous plots. See Fig.~\ref{fig-6} for model references. \label{fig-10}}
\end{figure}

\begin{figure}
\includegraphics[angle=0,scale=0.9]{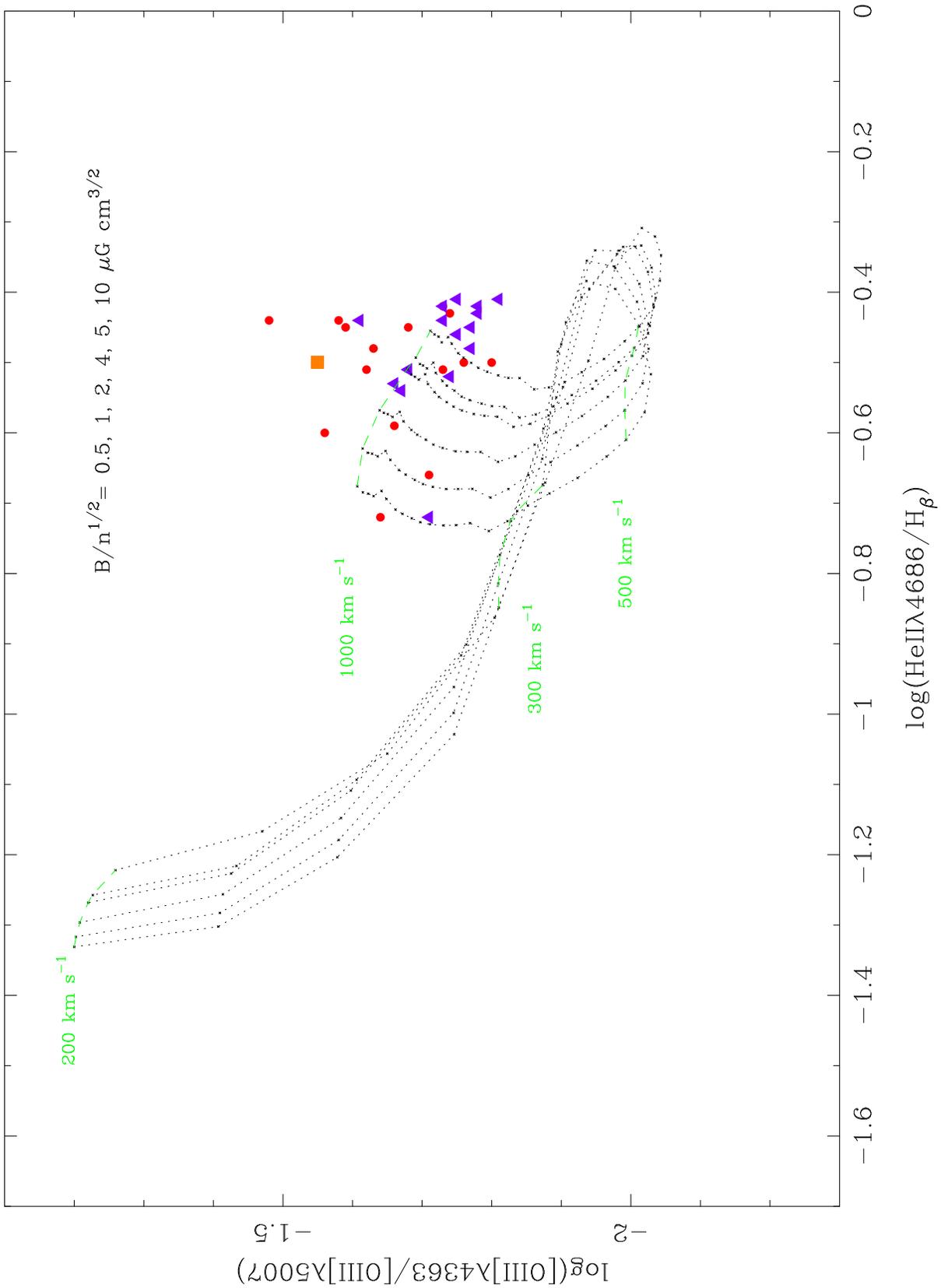}
\caption{The temperature-sensitive line ratio [O {\sc iii}]$\lambda4363$/[O {\sc iii}]$\lambda5007$ against He {\sc ii}/H$\beta$ as excitation axis. See Fig.~\ref{fig-6} for model references. \label{fig-11}}
\end{figure}

These models are also implemented through line-ratio diagnostic diagrams. We have drawn (Figures~\ref{fig-6}~to~\ref{fig-11}) the shock+precursor grids (dotted lines, one line per magnetic parameter\footnote{The magnetic parameter is defined as $B/n^{1/2}$, $B$ is the transverse magnetic field, and $n$ is the pre-shock gas density \citep{dop95}.}) from the most recent modelling of these phenomena by \citet{all08}. Models with solar abundance and pre-shock gas density of 100~cm$^{-3}$ were chosen. The only-shock grids (no precursor) lie outside the range of our observations \citep[compare the location of our data in Fig.~\ref{fig-6} with fig.~1 from][where both grids were plotted together]{dop95}. Shock velocities are also shown (dashed lines). Data are plotted as explained before.

We have used both [O {\sc iii}]$\lambda5007$/H$\beta$ and He {\sc ii}/H$\beta$ as excitation axis. These ratios, as well as the other pairs of lines employed, are close in wavelength to minimise the unknown reddening effects. In other cases we have used lines from the same species, to reduce the consequences that an unknown metallicity may have in our results \citep*{bal81}. As a general characteristic, our [O {\sc iii}]$\lambda5007$/H$\beta$ and He {\sc ii}/H$\beta$ values are higher than predicted by the model. In addition, our [O {\sc iii}]$\lambda5007$/H$\beta$ measurements are also much more constant. Anyway, the fit is good despite these minor inconsistencies, and the best fit is reached in Fig.~\ref{fig-7}, which employs those line-ratios.

The highly excited [Ne {\sc v}]$\lambda3424$ emission-line is used toghether with [O {\sc ii}]$\lambda3727$ in the next two plots (Fig.~\ref{fig-8} and \ref{fig-9}). The [Ne {\sc v}]$\lambda3424$ emission extends up to $\sim 5.5$~kpc away from the AGN; since $\sim 95$~eV photons are needed to produce this emission-line, it is very difficult for any photoionisation model to explain such an extended emission \citep{kom97}. In contrast, jet-cloud interaction may provide local energetic photons. As it is seen from the plots, these models also success in reproducing the relative intensity of [Ne {\sc v}]$\lambda3424$ with respect to the lower excited [O {\sc ii}]$\lambda3727$ emission-line.

The last two diagrams (Fig.~\ref{fig-10} and \ref{fig-11}) use the temperature-sensitive line-ratio [O {\sc iii}]$\lambda4363$/[O {\sc iii}]$\lambda5007$, which are particularly complicated to be reproduced by any kind of model. Since the ``temperature problem'' associated with this ratio \citep{dop95} is widely known, we do not draw any conclusion based upon these plots, but we show them because, despite the fact of having poorer fits, the data behave as in previous diagrams. 

Both high- and low-excitation emission lines are well fitted by the shock-ionisation model, although there is a clear underprediction regarding to [O {\sc iii}]$\lambda5007$ and He {\sc ii}. Furthermore, the extent to which the highest excited line in the spectrum is measured (i.e. [Ne {\sc v}$\lambda3424$]) requires a budget of energetic photons that can be locally created by this kind of fast radiative shocks. In fact, the behaviour of the ionisation parameter {\it U} is also explained by a local ionising field. Data tend to be located over the fastest (500~km~s$^{-1}$$<v<$1000~km~s$^{-1}$) shocks with no exception. It is worth highlighting that the disturbed [O {\sc iii}]$\lambda5007$ components reported on the velocity field move with velocities in the same range as those predicted by the shock-ionisation model. This is a strongly remarkable point since it is independent evidence to support the shock-ionisation idea as the main source of the ionisation of the gas.

Certainly the most spectacular interaction between the radio jets and the IGM is observed at the edges of the lobes, where the shock waves driven by the jet give raise to the bright, highly overpressured hotspots. The effects of shock waves in the way toward the hotspots have been also detected in the optical-emitting gas. An impressive example of both jet-cloud and lobe-cloud interaction was reported in 3C~299 \citep{fei99}, where the EELR overlaps the radio lobe. However, a turning point was then achieved by \citet{sol01} who showed that, despite the large distance in-between the EELR and the radio lobes in their sample, the EELR's gas undergoes jet-induced shock-related disturbing processes. In these latter cases the key evidence came from the strongly disturbed kinematical patterns observed within the gas. The same kind of disturbing motion that we find in 3C~381. 

Jet-driven shock waves are triggered near the AGN (near compared with the lobe-scale) since the jet is supersonic in these regions (Leahy; www.jb.man.ac.uk/atlas/dragns.html). Consequently, those shocks give raise to high-pressured regions within the clumpy EELR. Such a pressure-excess was reported by \citet{rob02}. They made use of their density measurements (we have used it at the beginning of this Section) to show that the EELR's gas is overpressured with respect to the X-ray halo. They claimed that jet-induced shocks might be playing an important role in the region, but no kinematical argument was then available to support this scenario. The velocity field we have shown in Section~\ref{kinematics} accounts for such an argument. 

We have shown that the jet-driven shock-ionising waves that are able to reproduce the observed line-ratio are also those that move with velocities entirely compatible with the high-velocity line-splitting shown by the velocity field. So, the kinematics of the region, its ionisation and also its pressure excess are now explained as consequences of the same process: the interaction of the radio-jet with the IGM.

%
%

\section{Conclusions}

From the study of the long-slit Gemini/GMOS spectra over the extended emission-line region found in the HST imaging of 3C~381, and the state-of-the-art modelling for the energetic balance for these structures, we find the following results:
 
- The nebular region extends up to 12 arcsec ($\sim 38$~kpc) and it is larger than the stellar component as seen in the HST data, which vanishes below the noise at 1.7 arcsec ($\sim 6$~kpc) from the nucleus.

- The kinematics of the EELR shows features of disturbing processes, where high velocity gas components are clearly found not only near the central region but also towards the outskirts of the EELR. Differences up to 600~km~s$^{-1}$ in velocity are observed within the nebular structure.  

- Line-ratios have been used to compare theoretical predictions with our data and we found that the shock-ionisation model with a precursor component predicts the observed line ratios. The fastest shocks (500~km~s$^{-1}<v<$~1000~km~s$^{-1}$) are capable of reaching the extreme values of our data; there are variations amongst diagrams, but there is no contradiction between any line ratio.

- Shock-ionisation models are likely to be the dominant ionising process taking place within the EELR. In addition to the diagnostic diagrams that reproduce the observed line-intensities with shock velocities coincident with what the velocity field shows, the local ionising photons created by the shocks themselves provide a plausible explanation for both the existence of very extended high-excitation lines and the increment of {\it U} despite the increasing distance.

\bsp
%
%

\section*{Acknowledgments}

We thank Dr. M. Allen for sending us his modelling, which was very helpful for this paper. Also, we want to thank to the Support Staff of the Gemini Observatory, and to the referee for his/her useful comments.
%
%


\newpage


\begin{table}
\caption{Log of HST observations}
\label{tbl-hst}
\begin{tabular}{ccccc}
\hline\hline
Camera & Date & Exp.Time & Filter Name & Exposures \\
\hline
WFPC2 & 1995 Feb 11 & 140s & F702W & 2 \\
 & 1995 Aug 13 & 300s & FR533N33 & 2 \\
\hline
\end{tabular}

\end{table}
%

\begin{table}
\caption{Log of Gemini observation}
\label{tbl-gem}
\begin{tabular}{ccccc}
\hline
\hline
Instrument & Slit & Grating & Date & Exp.Time \\
\hline
GMOS & 0.5 arcsec & B600\_G5303 & 2005 Apr 7 & 2400s \\
\hline
\end{tabular}

\end{table}
%

\begin{table}
\begin{minipage}{140mm}
\caption{Set of identified emission-lines in 3C 381 spectrum.}
\label{tbl-lines}
\begin{tabular}{lccccc}
\hline\hline
Line\footnote{Wavelengths in Angstroms} & \multicolumn{1}{c}{Flux 3.2 arcsec NW} & \multicolumn{1}{c}{Flux 1.6 arcsec NW} & \multicolumn{1}{c}{Flux GC\footnote{Galactic centre}} & \multicolumn{1}{c}{Flux 1.6 arcsec SE} & \multicolumn{1}{c}{Flux 3.2 arcsec SE} \\
\hline
[Ne {\sc iii}]$\lambda3345$ & - & 0.18 & 0.47 & - & - \\

[Ne {\sc v}]$\lambda3424$ & - & 0.81 & 1.05 & 0.57 & - \\

[O {\sc ii}]$\lambda3727$ & 9.93 & 4.28 & 2.58 & 3.01 & 4.91 \\

[Ne {\sc iii}]$\lambda3869$ & - & 1.47 & 1.19 & 1.16 & - \\

H$\delta$ (4100) & - & 0.22 & 0.19 & 0.19 & - \\

H$\gamma$ (4340) & - & 0.54 & 0.47 & 0.52 & - \\

[O {\sc iii}]$\lambda4363$ & - & 0.27 & 0.38 & 0.29 & - \\

He {\sc ii} (4686) & - & 0.26 & 0.31 & 0.29 & - \\

[Ar {\sc iv}]$\lambda4711$\footnote{Possible contamination from [Ne {\sc iv}]$\lambda4718$} & - & 0.04 & - & 0.11 & - \\

[Ar {\sc iv}]$\lambda4740$ & - & - & 0.07 & - & - \\

H$\beta$ (4861) & 1 & 1 & 1 & 1 & 1 \\

[O {\sc iii}]$\lambda4959$ & 4.24 & 4.1 & 4.36 & 4.53 & 4.33 \\

[O {\sc iii}]$\lambda5007$ & 13.86 & 12.42 & 13.32 & 13.57 & 12.57 \\

\hline
\end{tabular}
\end{minipage}
\end{table}
%

\newpage

%

\newpage
%

\newpage
%

\newpage
%

\newpage
%
\newpage
\newpage
\newpage
\newpage
\newpage
\newpage

\label{lastpage}

\end{document}